\documentclass{article}
\usepackage{spconf,amsmath,amsfonts,mathtools,bbm}
\usepackage{graphicx}
\usepackage{hyperref,bm}
\usepackage {spverbatim}
\ninept

\title{ 
Audio Signal Enhancement with Learning from Positive and Unlabelled Data
}
\name{Nobutaka Ito and Masashi Sugiyama}
\address{The University of Tokyo, Japan}
\begin{document}
%
\maketitle
\begin{abstract}\vspace{-2mm}
Supervised learning is a mainstream approach to audio signal enhancement
(SE) and requires parallel training data consisting of both
noisy signals and the corresponding clean signals. Such 
data can only be synthesised and are mismatched with real
data, which can result in poor performance {on real data}. {Moreover, clean signals may be inaccessible in certain scenarios, which renders this conventional approach infeasible.
}
Here we explore SE using
non-parallel training data consisting of noisy signals and noise, which can be easily recorded. We define the positive (P) and
the negative (N) classes as signal {inactivity} and {activity}, respectively.
We observe that the spectrogram patches of noise clips can be used as
P data and those of noisy signal clips as unlabelled data. 
Thus, 
 {\it learning from positive and unlabelled data} enables
 a convolutional neural network to learn to classify each spectrogram patch as P or N {to enable} SE.
\end{abstract}
\begin{keywords}
Audio signal enhancement, convolutional neural networks, learning from positive and unlabelled data, weakly supervised learning.
\end{keywords}

\vspace{-4mm}\section{Introduction}\vspace{-3mm}
\label{sec:intro}


\textbf{Audio signal enhancement. }This paper deals with audio signal enhancement (SE),\footnote{We use the term ``SE'' in the sense of ``noise suppression''.} the task of extracting a specific class of sounds (called a {\it signal}) while suppressing the other classes of sounds
 (called {\it noise}) from the mixture of them (called a {\it noisy signal}).
 Applications of SE include automatic speech recognition (ASR), music information retrieval, and sound event detection.
In this paper, we focus on single-channel SE, which 
{ applies even in situations where only one microphone is available.}
 See, e.g.,~\cite{Wang2020} for a multichannel SE method.

\begin{figure}
\centering
\includegraphics[width=\columnwidth]{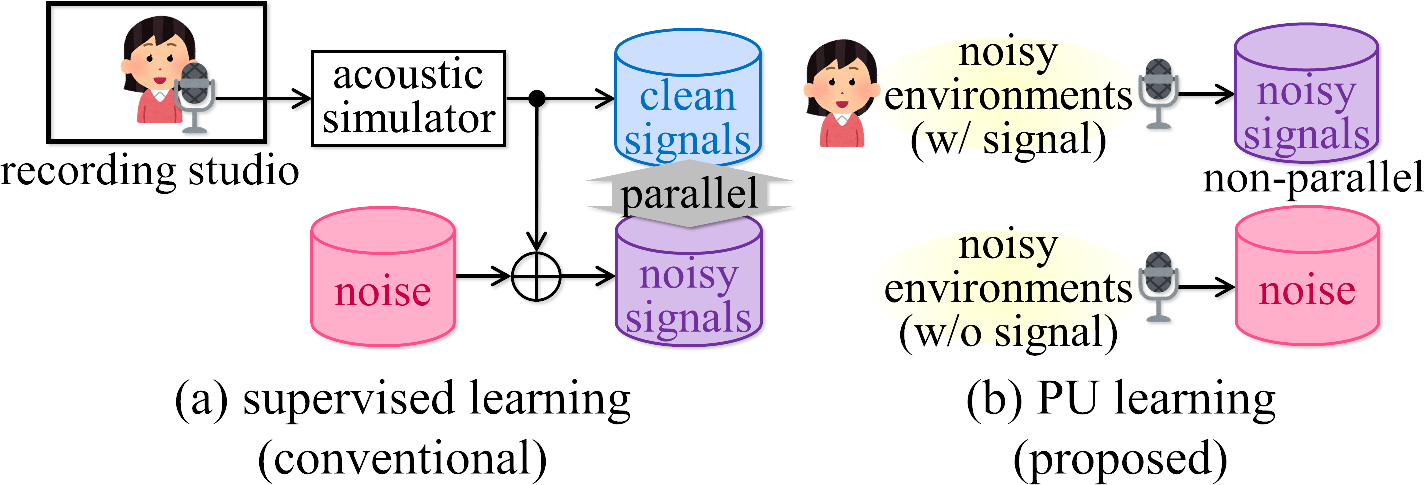}\vspace{-2mm}
\caption{Training data collection methods in conventional \& proposed approaches to SE. }\vspace{-2mm}
\label{fig:pipeline}
\end{figure}

\textbf{Supervised learning} has become a mainstream approach to SE, where
 an SE model such as a deep neural network (DNN) is trained 
 to predict the clean (i.e., noise-free) signal or a mask for SE~\cite{Erdogan2015,Tan2019,Hu2020}. 
The approach requires parallel training data consisting of both noisy signals and the corresponding clean signals.
As it is impossible to record such parallel data due to the crosstalk, a common practice is to synthesise them.
Fig.~\ref{fig:pipeline}(a) illustrates the data synthesis pipeline using speech enhancement as an example.
First, clean signals (e.g., clean speech utterances) are recorded in a {quiet} environment such as {a recording studio}. 
Then, these signals are {convolved with} room impulse responses, which are simulated by an acoustic simulator such as the image method~\cite{Allen1979}.
Finally, noisy signals are synthesised by numerically adding a clean signal and noise.


\textbf{Issues. }
There are fundamental issues in the above data collection method for supervised learning.
First, the synthetic training data
are inevitably mismatched with the distribution of real data that we wish to apply SE to.
This is because it is difficult to realistically simulate the characteristics of the sound source, the room sound field, and the recording device.
This can result in poor SE performance {on real data}, if the mismatch is severe.
Second, {clean signals may be inaccessible in certain scenarios, which renders this conventional approach infeasible.
For instance, clean signals may be inaccessible in the enhancement of music signals~\cite{Stoller2018GAN} or certain sound events like bird songs~\cite{Stowell2015,Denton2022}. }


\textbf{Motivation. }
The above issues motivate us to explore SE using
non-parallel {(i.e., unpaired)} training data consisting of noisy signals and noise (see Fig. 1(b)).
Note that noisy signals and noise
can be easily recorded in noisy environments, 
when the signal is active and inactive, respectively.


\textbf{Learning from positive and unlabelled data for audio signal enhancement. }
Here we propose {\it PULSE (PU learning for SE)}, an SE method using
non-parallel training data consisting of noisy signals and noise.
PULSE is based on {\it learning from positive and unlabelled data (PU learning)}~\cite{duPlessis2014,duPlessis2015,Kiryo2017,Sugiyama2022}, 
 a methodology for weakly supervised learning from only positive (P) and unlabelled (U) data without negative (N) data.
Let us define the P and the N classes as signal
{inactivity} and {activity}, respectively.\footnote{We could define the P class as signal {activity} but we do it the other way around to be consistent with PU learning.} 
The spectrogram patches of noise clips do not contain the signal by definition
and thus can be used as
P data.
On the other hand, those of noisy signal clips can be either P or N, depending on whether the signal is active or not in those spectrogram patches. Thus, they are used as U data.\footnote{
For a noisy signal clip, we know that there exists at least one spectrogram patch containing the signal. 
Here we {assume this constraint is satisfied.}}
These  P and U
data enable a convolutional neural network (CNN)  to learn to classify each
spectrogram patch as P or N through PU learning.
The trained CNN enables us to obtain a time--frequency (TF) dependent weight called a mask, 
which enables {TF components dominated by noise to be suppressed for SE.}
We also propose a {\it weighted sigmoid loss}, a new loss appropriate for SE, which uses the magnitude spectrogram as a weight in
 the sigmoid loss~\cite{Kiryo2017}.
The proposed loss turns out to be closely related to the signal approximation loss~\cite{Wang2018}
and empirically gave a substantially better speech enhancement performance than the original unweighted sigmoid loss~\cite{Kiryo2017}.
To
the best of our knowledge, PULSE is the first application of PU learning
to SE.

\vspace{-4mm}\section{Relation to prior work}\vspace{-3mm}

\textbf{Parallel training data without clean signals.} Like PULSE, {\it mixture invariant training (MixIT)}~\cite{Wisdom2020}\footnote{In \cite{Wisdom2020}, methods for source separation and SE were proposed and here we focus on the latter.} uses noisy signals and noise for training. {During training}, the sum of a noisy signal and noise is given as an input and a DNN has to separate it into an enhanced signal (i.e., an estimate of the clean signal) and two noise estimates. To enable training without ground truth clean signals, MixIT assumes that the sum of the enhanced signal and one of the noise estimates approximates the original noisy signal and the other noise estimate approximates the original noise. 
This approximation error is used as the loss, which can be computed without clean signals.
Note that there are two possibilities as to which of the noise estimates corresponds to the original noisy signal and it is unknown which is correct. To resolve this ambiguity, a mixture invariant loss similar to {the permutation invariant loss}~\cite{Kolbaek2017} is used. Unlike PULSE, however, MixIT requires parallel training data, where each training example is a triplet consisting of a noisy signal, noise, and their sum. As such parallel data cannot be recorded but can only be synthesised (i.e., the noisy signal and the noise are numerically added), {\it MixIT also suffers from the above mismatch issue.} For example, there may be mismatches in terms of the signal-to-noise ratio (SNR), the number of noise sources, etc.
A similar method was proposed in~\cite{Fujimura2021} and has the same mismatch issue.


\textbf{Non-parallel training data with clean signals.}
There are some methods that do not require parallel training data unlike many SE methods.
Such methods include generative modeling of clean signals~\cite{Smaragdis2007ICA,Bando2018,Subakan2018} and adversarial training of an SE network~\cite{Higuchi2017}. 
Unlike PULSE, these methods require clean signals for training and thus also 
suffer from the above issue of clean signals being often {inaccessible}.


\textbf{Unsupervised learning.}
Some SE methods require no supervision or even no training data at all based on unsupervised learning.
However, these methods require strong inductive biases based on domain knowledge (e.g., stationary noise)~\cite{Boll1979,Mohammadiha2013}
and are inapplicable when the assumptions made are violated.


\vspace{-3mm}\section{Preliminaries: PU learning}\vspace{-3mm}
\label{sec:prelim}
This section provides preliminaries on PU learning~\cite{duPlessis2014,duPlessis2015,Kiryo2017,Sugiyama2022}.
Let $\mathcal{X}^\mathrm{P}\coloneqq\{\mathbf{x}_{i}^\mathrm{P}\}_{i=1}^{n^\mathrm{P}}$ be P data and $\mathcal{X}^\mathrm{U}\coloneqq\{\mathbf{x}_{i}^\mathrm{U}\}_{i=1}^{n^\mathrm{U}}$ be U data, where $n^\mathrm{P}\coloneqq |\mathcal{X}^\mathrm{P}|$ and $n^\mathrm{U}\coloneqq |\mathcal{X}^\mathrm{U}|$ denote the number of P and U training examples, respectively. 
From these P and U data, a classifier has to learn the relationship between an input pattern $\mathbf{x}\in\mathbb{R}^d$ and a class label $y\in \{+1, -1\}$ so that it can predict the correct class label for an unseen pattern.
Here, $d$ is the feature dimension.
Let us assume that the P data $\mathcal{X}^\mathrm{P}$ are independent and identically distributed (i.i.d.) and follow an unknown class-conditional density $p(\mathbf{x}\hspace{-0.5mm}\mid\hspace{-0.5mm} y=+1)$. Let us also assume that the U data $\mathcal{X}^\mathrm{U}$ are i.i.d.\,\,and follow an unknown marginal density $p(\mathbf{x})$.
%

 Let $f_{\bm{\theta}}: \mathbb{R}^d\rightarrow \mathbb{R}$ be a classifier parametrised by $\bm{\theta}$ and $\widehat{y}\coloneqq\mathrm{sgn}( f_{\bm{\theta}}(\mathbf{x}))$ be the predicted label for $\mathbf{x}$, where $\mathrm{sgn}$ is the sign function
 \begin{align}
 \mathrm{sgn}({t})\coloneqq\begin{cases}
 +1&({t}\geq 0),\\
 -1&({t}< 0).
 \end{cases}
 \end{align}
 Let us define a {\it loss} $\ell(\mathbf{x},y,\bm{\theta})$, a non-negative function that measures the 
 deviation of the classifier $f_{\bm{\theta}}$ from a data point 
 $(\mathbf{x},y)$. 
 {Note that $\ell(\mathbf{x},y,\bm{\theta})$ can be any non-negative loss without any other constraints. }
 An example is the sigmoid loss~\cite{Kiryo2017}
 \begin{align}
 \ell(\mathbf{x},y,\bm{\theta})
 &=\sigma(-y f_{\bm{\theta}}(\mathbf{x})),\label{eq:sigloss}
 \end{align}
where $\sigma(m)\coloneqq (1+e^{-m})^{-1}$ is the logistic sigmoid function.
{It is worth noting that (\ref{eq:sigloss}) is bounded unlike many other common losses (e.g., the cross-entropy).}
 Let us define the {\it risk}, also known as (a.k.a.) the {\it generalisation error}, as
%
\begin{align}
R(\bm{\theta})\coloneqq\mathbb{E}_{p(\mathbf{x},y)}[\ell(\mathbf{x},y,\bm{\theta})],
\label{eq:risk0}
\end{align}
which is the expectation of the loss with respect to (w.r.t.)\,\,an unknown joint data distribution $p(\mathbf{x},y)$.
As $p(\mathbf{x},y)$ is unknown, the expectation in (\ref{eq:risk0}) cannot be computed in practice. Therefore, we compute an {\it empirical risk} (a.k.a. a {\it training error}) $\widehat{R}(\bm{\theta})\approx {R}(\bm{\theta})$ by replacing expectation by empirical averaging over training data and obtain $\bm{\theta}$ by solving the minimisation of $\widehat{R}(\bm{\theta})$.
This formulation is called {\it empirical risk minimisation}.



In supervised learning, an empirical risk is easily obtained as $R(\bm{\theta})\approx\frac{1}{n}\sum_{i=1}^n \ell(\mathbf{x}_{i},y_{i},\bm{\theta})$, where $\{(\mathbf{x}_{i},y_i)\}_{i=1}^n\sim p(\mathbf{x},y)$  are labelled training data.
 In contrast, in PU learning, only P and U data are given {as training data} without N data. It may appear to be impossible to compute an empirical risk using only such data, but this turns out to be possible as follows.

Note first that
\begin{align}
p(\mathbf{x},y)=\begin{cases}
\pi p(\mathbf{x}\mid y=+1)&(y=+1),\\
(1-\pi) p(\mathbf{x}\mid y=-1)&(y=-1),
\end{cases}
\label{eq:joint}
\end{align}
where $\pi\coloneqq {p}(y=+1)$ is the class prior for the P class and assumed to be given in this paper.
See~\cite{Sugiyama2022,Jain2016,duPlessis2017,Yao2022} for some methods for estimating $\pi$ from only P and U data.
Using (\ref{eq:joint}),
(\ref{eq:risk0}) is rewritten as
\begin{align}
R(\bm{\theta})
&=\pi\mathbb{E}_{p(\mathbf{x}\mid y=+1)}[\ell (\mathbf{x},+1,\bm{\theta})]\label{eq:risk1024}\\
&\phantom{=}+(1-\pi)\mathbb{E}_{p(\mathbf{x}\mid y=-1)}[\ell (\mathbf{x},-1,\bm{\theta})].\notag
\end{align}
Since $p(\mathbf{x})=\pi p(\mathbf{x}\mid y=+1)+(1-\pi) p(\mathbf{x}\mid y=-1)$,
(\ref{eq:risk1024}) is further rewritten as
\begin{align}
R(\bm{\theta})
&=\pi\mathbb{E}_{p(\mathbf{x}\mid y=+1)}[\ell (\mathbf{x},+1,\bm{\theta})]\label{eq:risk2048}\\
&\phantom{=}+\mathbb{E}_{p(\mathbf{x})}[\ell (\mathbf{x},-1,\bm{\theta})]-\pi\mathbb{E}_{p(\mathbf{x}\mid y=+1)}[\ell (\mathbf{x},-1,\bm{\theta})].\notag
\end{align}
Therefore, we obtain an empirical risk using only P and U data, $\widehat{R}_\mathrm{PU}(\bm{\theta})\approx R(\bm{\theta})$, as follows~\cite{duPlessis2015}:
\begin{align}
\widehat{R}_\mathrm{PU}(\bm{\theta})\coloneqq &\frac{\pi}{|\mathcal{X}^\mathrm{P}|}\sum_{\mathbf{x}\in\mathcal{X}^\mathrm{P}}\ell (\mathbf{x},+1,\bm{\theta})\label{eq:RPU}\\
&+\frac{1}{|\mathcal{X}^\mathrm{U}|}\sum_{\mathbf{x}\in\mathcal{X}^\mathrm{U}}\ell (\mathbf{x},-1,\bm{\theta})-\frac{\pi}{|\mathcal{X}^\mathrm{P}|}\sum_{\mathbf{x}\in\mathcal{X}^\mathrm{P}}\ell (\mathbf{x},-1,\bm{\theta}).\notag
\end{align}

While
$\mathbb{E}_{p(\mathbf{x})}[\ell (\mathbf{x},-1,\bm{\theta})]-\pi\mathbb{E}_{p(\mathbf{x}\mid y=+1)}[\ell (\mathbf{x},-1,\bm{\theta})]$ in (\ref{eq:risk2048}) is non-negative for a non-negative loss $\ell$, its empirical approximation in (\ref{eq:RPU}) can be negative. This is especially significant for an unbounded loss (e.g., the cross-entropy) or a flexible model (e.g., a DNN).
To remedy this, the following non-negative empirical risk $\widehat{R}_\mathrm{nnPU}(\bm{\theta})\approx R(\bm{\theta})$ was proposed~\cite{Kiryo2017}:
\begin{align}
\widehat{R}_\mathrm{nnPU}(\bm{\theta})\coloneqq &\frac{\pi}{|\mathcal{X}^\mathrm{P}|}\sum_{\mathbf{x}\in\mathcal{X}^\mathrm{P}}\ell (\mathbf{x},+1,\bm{\theta})+\Biggl(\frac{1}{|\mathcal{X}^\mathrm{U}|}\sum_{\mathbf{x}\in\mathcal{X}^\mathrm{U}}\ell (\mathbf{x},-1,\bm{\theta})\notag\\
&-\frac{\pi}{|\mathcal{X}^\mathrm{P}|}\sum_{\mathbf{x}\in\mathcal{X}^\mathrm{P}}\ell (\mathbf{x},-1,\bm{\theta})\Biggr)_+,\label{eq:RnnPU}
\end{align}
where `nn' stands for `non-negative' and 
\begin{align}
(x)_+\coloneqq\begin{cases}
x,&(x\geq 0),\\
0,&(x <0).
\end{cases}
\end{align}

\begin{figure}
\centering
\includegraphics[width=\columnwidth]{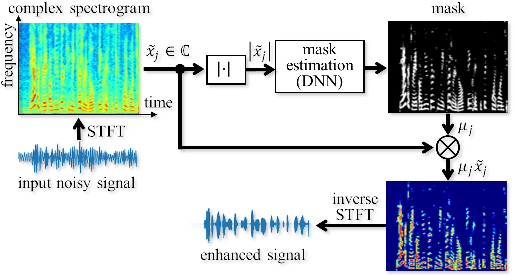}\vspace{-4mm}
\caption{Pipeline of masking-based SE.
}\vspace{-5mm}
\label{fig:MIL}
\end{figure}

\vspace{-4mm}\section{PULSE: PU learning for SE}\vspace{-3mm}
\label{sec:PULSE}

\textbf{Masking-based SE. }
In this paper, we employ a masking-based approach to SE, where we use a DNN to estimate a mask in the short-time Fourier transform (STFT) domain (Fig.~\ref{fig:MIL}). 
In this approach,
 an input noisy signal clip in the time domain is first transformed into the TF domain by the STFT. Let the resulting STFT-domain representation (i.e., the complex spectrogram) be $\widetilde{x}_j\in\mathbb{C}$, where $j$ is the TF component index. Then, 
 we compute a magnitude spectrogram $|\widetilde{x}_j|$ by taking the absolute value.
A DNN is given the magnitude spectrogram and 
estimates
a mask $\mu_j$.
An enhanced signal  is obtained by elementwise multiplication (i.e., masking) $\mu_j\widetilde{x}_j$,
which suppresses TF components dominated by noise. Finally,
the enhanced signal is transformed back into the time domain by the inverse STFT.

\textbf{Motivation. }
In this approach, it is crucial to train the DNN so that the mask can be estimated properly.
If we were given parallel training data consisting of both noisy signals and the corresponding clean signals, we could do so by supervised learning in a straightforward way.
However, as we already mentioned in Sec.~\ref{sec:intro}, such a training methodology has fundamental issues, which motivated us to develop PULSE.

\begin{figure}
\centering
\includegraphics[width=\columnwidth]{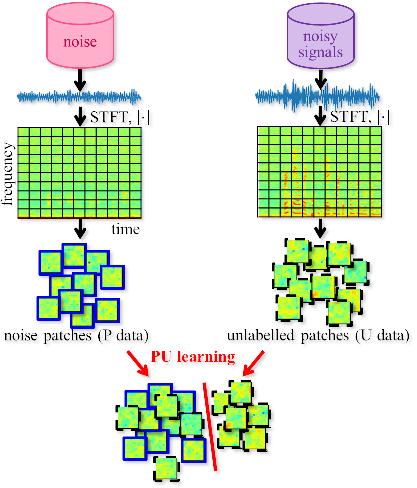}\vspace{-3mm}
\caption{Pipeline of PULSE.}\vspace{-5mm}
\label{fig:PULSE}
\end{figure}

\textbf{Pipeline of PULSE} is illustrated in
Fig.~\ref{fig:PULSE}.
The training data consist of  noisy signal clips and noise clips.
We first compute a magnitude spectrogram of 
each clip  by applying the STFT and then taking the absolute value.
Then, we crop a rectangular spectrogram patch centred at each TF point as the input feature.
Let us define a TF component as P and N
if the signal is {inactive} and {active} in the corresponding spectrogram patch, respectively.
Each TF {component} of a noise
clip is P.
On the other hand,
each TF {component} of a noisy signal clip 
can be either P or N and is thus treated as U.
Thus, $\mathcal{X}^\mathrm{P}$ and $\mathcal{X}^\mathrm{U}$ consist of the spectrogram patches of 
noise clips and those of noisy signal clips, respectively.
These P and U data are used to train
a CNN to classify each TF {component} as P or N by PU learning described in Sec.~\ref{sec:prelim}.
During testing, the mask $\mu_j$ is obtained 
by 
\begin{align}
\mu_j\leftarrow \begin{cases}
1&(\widehat{y}_j=-1),\\
0&(\widehat{y}_j=+1).
\end{cases}
\label{eq:mu}
\end{align}
Here,
$\widehat{y}_j\coloneqq \mathrm{sgn}(f_{\widehat{\bm{\theta}}}(\mathbf{x}_j))$ is the predicted label of the $j$th TF component,
where $\mathbf{x}_j$ is the corresponding spectrogram patch and $\widehat{\bm{\theta}}$ is the trained parameters.
This mask retains the TF components classified as N 
and removes those classified as P.

\textbf{Architecture. } 
The classifier $f_{\bm{\theta}}$ is modelled by the following 11-layer CNN:
Compress($1/15$)-Conv2d(1, 8, 3)-Conv2d(8, 8, 3)-Conv2d(8, 16, 3)-Conv2d(16, 16, 3)-Conv2d(16, 32, 3)-Conv2d(32, 32, 3)-Conv2d(32, 64, 3)-Conv2d(64, 64, 3)-Conv2d(64, 128, 1)-Conv2d(128, 128, 1)-Conv2d(128, 1, 1). 
Here, Compress($\alpha$) is a
power-law compression layer that applies the non-linear function
$(\cdot)^\alpha$ elementwise.
Conv2d($C_\mathrm{in}$, $C_\mathrm{out}$, $K$) is a two-dimensional convolutional layer with $C_\mathrm{in}$ input channels,
$C_\mathrm{out}$ output channels, a kernel size of $K\times K$, a stride of $(1,1)$, and no padding.
All but the last convolutional layer are followed by a
rectified linear unit (ReLU) and then a
dropout layer with a {dropout} rate of 0.2.
The size of the input spectrogram patch is set to that of the receptive field of the network (i.e., $17\times 17$)
so that the CNN output size is $1\times 1$.

\textbf{Loss. }
It is crucial to design the loss $\ell(\mathbf{x},y,\bm{\theta})$ properly to obtain a good SE performance by PULSE.
It measures the deviation of
the classifier $f_{\bm{\theta}}$ from  $(\mathbf{x},y)$, where
$\mathbf{x}$ is a spectrogram patch and $y$ is the corresponding label.
{Commonly used losses in PU learning, such as the sigmoid loss (\ref{eq:sigloss}) or the cross-entropy, assign uniform weights to all TF components. 
}
As {the classification accuracy of}
the TF component with a larger magnitude is more significant in SE, we
 introduce the magnitude spectrogram $w(\mathbf{x})\coloneqq |\widetilde{x}|$ as a weight in (\ref{eq:sigloss}). Specifically, our loss is given by
\begin{align}
 \ell(\mathbf{x},y,\bm{\theta})=w(\mathbf{x})\sigma(-y f_{\bm{\theta}}(\mathbf{x})),\label{eq:wsigloss}
 \end{align}
 which we call a
{\it weighted sigmoid loss}.\footnote{To give alternative interpretations of (\ref{eq:wsigloss}), note that it can be rewritten as
$
 \ell(\mathbf{x},y,\bm{\theta})=|\frac{1}{2}(y+1)\widetilde{x}-(\sigma\circ f_{\bm{\theta}})(\mathbf{x})\widetilde{x}|
$, where $\circ$ is function composition.
Here, $\frac{1}{2}(y+1)$ can be interpreted as a ground truth binary mask for extracting noise as it retains the P TF components and removes the
 N TF components.
On the other hand, $(\sigma\circ f_{\bm{\theta}})(\mathbf{x})$ can be regarded as an estimated soft mask for extracting noise as
it approaches $1$ when $f_{\bm{\theta}}(\mathbf{x})$ approaches $+\infty$
 and approaches $0$ when $f_{\bm{\theta}}(\mathbf{x})$ approaches $-\infty$.
  Therefore, our loss is the absolute error between a noise estimate $\frac{1}{2}(y+1)\widetilde{x}$
  using the ground truth binary mask $\frac{1}{2}(y+1)$
  and a noise estimate $(\sigma\circ f_{\bm{\theta}})(\mathbf{x})\widetilde{x}$ using the estimated soft mask $(\sigma\circ f_{\bm{\theta}})(\mathbf{x})$.
  Our loss can also be seen as the signal approximation loss~\cite{Wang2018} with the following modifications. First, our loss is defined w.r.t.\,noise instead of the signal. Second,
  the target in our loss is the noise estimate using the ground truth mask, $\frac{1}{2}(y+1)\widetilde{x}$, instead of the ground truth noise itself. 
  Third, our loss  uses
  the $1$-norm  instead of the $2$-norm.
}

\textbf{Clip-wise processing.} 
The above patch-wise processing is computationally inefficient as it repeats some operations common to neighbouring spectrogram patches.
A more efficient clip-wise processing consists in simply applying the above CNN to entire magnitude spectrograms
instead of spectrogram patches, similar to~\cite{Sermanet2014,Redmon2016}.
In this case, the `same' padding is applied to each convolutional layer to preserve the dimensions of the spectrogram.

%

\vspace{-4mm}\section{Speech Enhancement Experiment}\vspace{-3mm}
We conducted a preliminary experiment of speech enhancement using PyTorch
as a proof of concept
to confirm 
the feasibility of SE using non-parallel data through PULSE.\footnote{The source code is available at https:\slash\slash{}github.com\slash{}nobutaka-ito\slash{}pulse.}

\textbf{Methods.}
We compared the following methods:
\begin{itemize}
\item \textbf{PULSE.} 
The 11-layer CNN (clip-wise processing) in Sec.~\ref{sec:PULSE} was trained by PU learning with the non-negative empirical risk (\ref{eq:RnnPU})
(specifically Algorithm~1 of~\cite{Kiryo2017}). {The weighted sigmoid loss (\ref{eq:wsigloss}) was used.
The class prior was set to $\pi=0.7$, which was tuned on the validation set.}
During testing, {a binary} mask was obtained by (\ref{eq:mu}).
\item \textbf{Supervised learning.} The same CNN architecture was used except that the kernel size was $3\times 3$ in all convolutional layers. The CNN was trained by supervised learning with the signal approximation loss in the STFT domain~\cite{Wang2018}. {A soft} mask was obtained by applying the sigmoid activation to the CNN output.
{\item \textbf{MixIT~\cite{Wisdom2020}: }The same architecture as in \textbf{supervised learning} was used except that there were three output channels corresponding to the enhanced signal and two noise estimates. The CNN was trained by solving the minimisation of
a mixture invariant loss (see~\cite{Wisdom2020} for details) based on the signal approximation loss in the STFT domain~\cite{Wang2018}. 
{Soft} masks for the enhanced signal and two noise estimates were obtained by applying the sigmoid activation to the CNN output.}
\end{itemize}
In all methods, the frame length and the hop for the STFT were set to 1024 samples (64\,ms) and 256 samples (16\,ms), respectively,
and the hamming window was used.

\textbf{Data.} 
We  focused on synthetic data. This is because most evaluation metrics for speech enhancement performance,
 including the scale-invariant SNR (SI-SNR)~\cite{LeRoux2019}, require parallel data, which can only be synthesised. We will conduct an evaluation on real data w.r.t.\,\,the ASR accuracy or a non-intrusive metric, such as DNSMOS~\cite{Reddy2021}, in future work.
We prepared a speech enhancement dataset using speech from TIMIT~\cite{TIMIT} and noise from DEMAND~\cite{Thiemann2013}.
We used the training set of TIMIT to create our training and validation sets 
and the test set of TIMIT to create our test set.
We used noise recordings from DEMAND in the following environments, which we found contained little speech: DKITCHEN, DLIVING, DWASHING, NFIELD, NRIVER, OHALLWAY, OOFFICE, STRAFFIC, and TCAR.
Each noise recording was divided into halves,
one for the training and the validation sets and the other for the test set.
The training set for \textbf{PULSE} consisted of 4019 noisy speech clips and 4019 noise clips (3.49\,h { each}).
Throughout this experiment,
 all clips were 3.125\,s long and sampled at 16\,kHz.
The training set for \textbf{supervised learning} consisted of 4019 noisy speech clips along with the corresponding clean speech clips.
{The training set for \textbf{MixIT} consisted of 4019 noisy speech clips,
4019 noise clips, and the corresponding 4019 mixture (i.e., noisy speech plus noise) clips.}
 In all methods, the validation/test set consisted of 601/1680 noisy speech clips (0.52\,h/1.46\,h)
along with the corresponding clean speech clips. 
Each noise clip above was a random excerpt from DEMAND;
Each noisy speech clip  was generated by adding a TIMIT clip and a random excerpt from DEMAND at an SNR sampled uniformly from the interval $[-5,10]$\,dB.

\textbf{Training.} 
Data-parallel distributed training was performed on 16 NVIDIA A100 GPUs for 400 epochs with the Adam optimiser~\cite{Kingma2015}, a batch size per GPU of 16, and a learning rate of 0.0018/0.0032/0.00055 (\textbf{PULSE}/\textbf{supervised learning}/\textbf{MixIT}).

\textbf{Metric.} 
We used the SI-SNR~\cite{LeRoux2019} as the evaluation metric for speech enhancement performance.
Let $\mathbf{s}$ be  the clean speech and $\widehat{\mathbf{s}}$ be an estimate of it, both in the time domain.
Note that $\widehat{\mathbf{s}}$ can be decomposed as $\widehat{\mathbf{s}}=\widehat{\mathbf{s}}_\parallel+\widehat{\mathbf{s}}_\perp$.
Here, $\widehat{\mathbf{s}}_\parallel\coloneqq \frac{\mathbf{s}^T\widehat{\mathbf{s}}}{\|\mathbf{s}\|^2}\mathbf{s}$
is the component parallel to $\mathbf{s}$
and 
 $\widehat{\mathbf{s}}_\perp\coloneqq \widehat{\mathbf{s}}-\widehat{\mathbf{s}}_\parallel$
 is that perpendicular to it, where $(\cdot)^T$ is transposition and $\|\cdot\|$ is the 2-norm.
The  SI-SNR is defined using the ratio of the squared norms of these components as follows:
\begin{align}
\text{SI-SNR}(\mathbf{s},\widehat{\mathbf{s}})\coloneqq 10\log_{10}\frac{\|\widehat{\mathbf{s}}_\parallel\|^2}{\|\widehat{\mathbf{s}}_\perp\|^2}=10\log_{10}\frac{\|a\mathbf{s}\|^2}{\|\widehat{\mathbf{s}}-a\mathbf{s}\|^2}
\end{align}
with $a\coloneqq \frac{\mathbf{s}^T\widehat{\mathbf{s}}}{\|\mathbf{s}\|^2}$.
It is convenient to define an SI-SNR improvement (SI-SNRi) as
the difference of $\text{SI-SNR}(\mathbf{s},\widehat{\mathbf{s}})$
with $\widehat{\mathbf{s}}$ being the enhanced speech and
$\text{SI-SNR}(\mathbf{s},\widehat{\mathbf{s}})$ with $\widehat{\mathbf{s}}$ being
the observed noisy speech.
Not only did we use the test SI-SNRi for performance evaluation,
we also tuned the hyperparameters based on the validation SI-SNRi.
A model checkpoint was saved at the end of each epoch and the one with the highest validation SI-SNRi was selected for evaluation on the test set.

{
\textbf{Results. }
Table~\ref{table:SI-SNRi} shows the SI-SNRi on the test set.
By using only non-parallel training data, which can be easily recorded, \textbf{PULSE}
 was able to give an SI-SNRi of as much as 14.62\,dB.
 This score was superior to 12.19\,dB for \textbf{MixIT} and
 close to 15.86\,dB for \textbf{supervised learning}.
 For ablation, we also evaluated PULSE with the original sigmoid loss (\ref{eq:sigloss}) instead of the weighted one (\ref{eq:wsigloss})
 and PULSE with the original empirical risk (\ref{eq:RPU}) instead of the non-negative one (\ref{eq:RnnPU}).
 Without (\ref{eq:wsigloss}) and (\ref{eq:RnnPU}), the SI-SNRi dropped to 9.30\,dB and 12.96\,dB, respectively, which shows the significance of the loss weighting and the non-negative empirical risk.
}

\begin{table}[h]

 \caption{SI-SNRi on the test set. The mean over five trials is shown along with
the standard deviation in the parentheses.}
 \label{table:SI-SNRi}
 \centering
  \begin{tabular}{lc}
   method&SI-SNRi (dB)\\\hline
   PULSE&14.62 \footnotesize{(0.20)}\\
   supervised learning&15.86 \footnotesize{(1.28)}\\
   MixIT~\cite{Wisdom2020}&12.19 \footnotesize{(4.50)}\\\hline
   PULSE w/o (\ref{eq:wsigloss})&9.30 \footnotesize{(0.70)}\\
   PULSE w/o (\ref{eq:RnnPU})&12.96 \footnotesize{(3.19)}
  \end{tabular}
\end{table}

\vspace{-4mm}\section{Conclusions}\vspace{-3mm}
We proposed PULSE, a PU-learning-based paradigm for training SE models {on}
non-parallel data consisting of noisy signals and noise.
The feasibility of PULSE was confirmed through a speech enhancement experiment.
The future work shall include exploring more sophisticated architectures, conducting  more extensive experiments, and applying PU learning to other audio tasks.

\vspace{2mm}
\textbf{Acknowledgment.}\ \ 
The authors were supported by the Institute for AI and Beyond, UTokyo.



\end{document}